\documentclass[a4paper]{llncs}

\SetMathAlphabet{\mathcal}{normal}{OMS}{xmdcmsy}{m}{n}
\usepackage{hyperref}
\hypersetup{%
  colorlinks=true,%
  citecolor=blue,%
  urlcolor=blue,%
  linkcolor=blue,%
}

\usepackage{isabelle}
\usepackage{stmaryrd}
\usepackage{tikz}
\usepackage{amssymb}
\usepackage{xspace}
\usepackage{ifthen}
\usepackage{amsmath}  
\usepackage{array} 
\usepackage[english]{babel}
\usepackage{cite}% smaller spaces after commas of cites

\newcommand{\rSC}[1]{\secref{#1}}
\renewcommand{\theenumi}{\roman{enumi}}

\makeatletter
\renewcommand{\p@enumii}{\theenumi--}
\makeatother

\newcommand{\rSub}[1]{(\ref{#1})}

\newcommand\TTTT{%
 \ensuremath{\textsf{T\kern-0.2em\raisebox{-0.3em}T\kern-0.2emT\kern-0.2em\raisebox{-0.3em}2}}%
}
\newcommand{\aprove}{{\sf APro\kern-0.1emVE}}

\newcommand\ceta{\textsf{C\kern-0.2exe\kern-0.5exT\kern-0.5exA}\xspace}
\newcommand\isafor{\textsf{Isa\kern-0.2exF\kern-0.2exo\kern-0.2exR}\xspace}

\newcommand\nats{\mathbb{N}}

\newcommand\FF{\mathcal{F}}

\newcommand\RR{\mathcal{R}}

\newcommand\VV{\mathcal{V}}

\newcommand\SN[1][]{\mathsf{SN}_{#1}}

\newcommand\rstep[1][\RR]{\to_{#1}}

\newcommand\defref[2][]{Def.~\ref{#2}\ifthenelse{\equal{#1}{}}{}{\parref{#1}}}

\newcommand\lemref[2][]{Lem.~\ref{#2}\ifthenelse{\equal{#1}{}}{}{\parref{#1}}}

\newcommand\parref[1]{{(\ref{#1})}}

\newcommand\secref[1]{Sec.~\ref{#1}}

\newcommand\thmref[2][]{Thm.~\ref{#2}\ifthenelse{\equal{#1}{}}{}{\parref{#1}}}

\let\OLDparagraph\paragraph
\renewcommand\paragraph[1]{\OLDparagraph{\sl#1\/}}

\renewcommand{\theenumi}{\textsf{\alph{enumi}}}
\begin{document}

\mainmatter

\title{Towards the Certification of Complexity Proofs\thanks{%
This research is supported by the Austrian Science Fund (FWF): P22767-N13.}}

\author{
%Christian Sternagel\inst{1} \and 
Ren\'e Thiemann%\inst{2}
}

\institute{
%  Japan Advanced Institute of Science and Technology, Japan
%\and
Institute of Computer Science, University of Innsbruck, Austria 
}

%------------------------------------------------------------------------
\maketitle

\begin{abstract}
We report on our formalization of matrix-interpretation in Isabelle/HOL.
Matrices are required to certify termination proofs and we wish to utilize
them for complexity proofs, too.
For the latter aim, only basic methods have already been integrated, and we discuss
some upcoming problems which arise when formalizing more complicated 
results on matrix-interpretations, which are based on Cayley-Hamilton's
theorem or joint-spectral radius theory.
\end{abstract}

%------------------------------------------------------------------------
\section{Introduction}

\isafor is a an \emph{Isa}belle/HOL \cite{Isabelle} \emph{Fo}rmalization of \emph{R}ewriting \cite{CeTA}. 
The initial aim in the development of \isafor was the certification of 
termination proofs of term rewrite systems (TRSs). Here, several important
techniques like recursive path orders, polynomial orders, matrix interpretations,
and dependency pairs have been formalized in a deep embedding. All these termination
techniques are accompanied with executable algorithms which guarantee the correct
application of these techniques in some termination proof that should be 
certified. The corresponding certifier \ceta (\emph{Ce}rtified 
\emph{T}ermination \emph{A}nalysis)
is obtained by invoking Isabelle's
code-generator \cite{codegen10} on these executable algorithms. 

In the mean time, most termination techniques that are applied in current
termination tools for TRSs can indeed be certified, and \isafor was extended
towards other interesting rewriting related 
topics like confluence, completion, and complexity analysis.

In the sequel, we will report on our formalization of complexity analysis,
where we will concentrate on one specific method: matrix 
interpretations for inferring polynomial complexity bounds.
To this end, we will shortly recapitulate some theory on term rewriting
and matrix interpretations
 in \rSC{basics}.  Our formalization of matrix interpretations
for termination proofs in presented in \rSC{matrix}. We discuss the extension
to complexity proofs in \rSC{complexity} where we also discuss some open problems.

All formalizations described in this paper are available from the AFP-entry \cite{matrixAFP} or from the
\isafor-library (\url{http://cl-informatik.uibk.ac.at/software/ceta}).

%------------------------------------------------------------------------
\section{Preliminaries}
\label{basics}

We assume  familiarity with term rewriting \cite{BN98}. Still, we
recall the most important notions that are used later on. A 
\emph{term}~$t$ over a set of \emph{variables}~$\VV$ and a set of
\emph{function symbols}~$\FF$ is
either a variable $x \in \VV$ or an \mbox{$n$-ary} function symbol~$f \in \FF$ applied to
$n$ argument terms $f(t_1, \ldots, t_n)$. We write $|t|$ for the size of a term.

A \emph{rewrite rule} is a pair of terms
$\ell \to r$ and a TRS~$\RR$ is a set of rewrite rules.
The \emph{rewrite relation (induced by $\RR$)} $\rstep$ is the closure
under substitutions and under contexts of $\RR$, i.e., $s \rstep t$ iff
there is a context~$C$, a rewrite rule $\ell\to r \in \RR$, and a substitution
$\sigma$ such that
$s = C[\ell\sigma]$ and $t = C[r\sigma]$. 
A TRS~$\RR$ is terminating, written $\SN(\RR)$, if there is no
infinite derivation $t_1 \to_\RR t_2 \to_\RR t_3 \to_\RR \dots$.

\newcommand{\dc}{\mathsf{dc}_\RR}
\newcommand{\tor}{\to_\RR}

For a finite and terminating TRS $\RR$, we its \emph{derivational 
complexity} $\dc : \nats \to \nats$ is defined as $\dc(n) = \max \{k \mid \exists t_1 \dots t_k. |t_1| \leq n \wedge t_1 \tor t_2 \tor \dots \tor t_k \}$.

\newcommand{\pl}{\oplus}
\newcommand{\ti}{\odot}
\newcommand{\one}{\underline{\mathsf{1}}}
\newcommand{\zero}{\underline{\mathsf{0}}}
\newcommand{\mono}{\mathsf{mono}}
\newcommand{\pos}{\mathsf{pos}}
\renewcommand{\max}{\mathsf{max}}
\newcommand{\mat}[1]{#1^{n \times n}}
\newcommand{\matsd}[1]{#1_{sd}^{n \times n}}
\newcommand{\A}{{\cal A}}
\newcommand{\sema}[1]{[\![#1]\!]_\alpha}
\newcommand{\semo}[1]{[\![#1]\!]}
\newcommand{\MM}{{\cal M}}
\newcommand{\OO}{{\cal O}}
\newcommand{\norm}[1]{||#1||}

One important termination technique is the usage of
well-founded monotone algebras. In this approach, it is
assumed that there is some algebra $(\A,\{f_\A\}_{f \in \FF})$,
where $\A$ is the universe and for each function symbol 
$f$ or arity $n$ we have an interpretation 
function $f_\A : \A^n \to \A$. Moreover, there is some well-founded order
$>$ on $\A$ and all $f_\A$ have to be monotone w.r.t.~$>$ in all their arguments.

Proving termination using well-founded monotone algebras can now be done
by demanding for all assignments $\alpha : \VV \to \A$ and all rules
$\ell \to r \in \RR$ that $\sema l > \sema r$. The reason is that then 
every rewrite step $s = C[\ell\sigma] \tor C[r\sigma] = t$ leads to a strict
decrease $\sema{s} > \sema{t}$ w.r.t.\ the well-founded order.
Polynomial orders \cite{CL87,L79} are a well-known instance of well-founded monotone
algebras where every $f_\A$ is
a polynomial, $\A = \nats$, and $>$ is the standard order on the naturals.

Despite proving termination, well-founded monotone algebras can also be used
for complexity analysis. Assume $>$ is the standard order on the naturals.
Then for any ground term $t$, its interpretation $\semo t$ is a bound on
the length of each derivation starting in $t$. Hence, 
if we can find a bound $b : \nats \to \nats$ such that $\semo t \leq b( |t| )$
for all terms $t$,
then $\dc(n) \leq b(n)$ and thus, $b$ is also a bound for the derivational
complexity.

Unfortunately, if one considers polynomial orders, then the bound $b$ can be double-exponential and this bound is tight \cite{HL89}.
Even for linear polynomial orders in general one can only infer an exponential
bound. Only for a very restricted class of polynomial interpretations 
(strongly linear interpretations \cite{Bon01}), one achieves a linear complexity bound.
So, when using polynomial interpretations we either have to impose severe restrictions to obtain a linear bound, or
without this restriction we can only guarantee non-polynomial bounds.

Luckily, it turned out that other well-founded monotone algebras are useful for
proving termination \cite{MatrixJAR} and complexity \cite{MSW08}: matrix-interpretations. Matrix interpretations
are similar to linear polynomial interpretations except that $\A$ is the
set of $n$-dimensional square matrices over some carrier.\footnote{In
\cite{MatrixJAR}, the universe consists of vectors, and the linear 
interpretations
take matrices as coefficients. 
However, in the formalization of \cite{SOFSEM10} and also in our
formalization of matrix interpretations it was easier to always use matrices.} 
To be more precise, every $f_\A$ is of the form $f_\A(x_1,\ldots,x_n) = 
M_{f,0} + M_{f,1}x_1 + \ldots + M_{f,n}x_n$ and matrices are compared 
by demanding a strict decrease in the upper-left entry, and a weak decrease
in all remaining entries.
To ease presentation
we here assume that we have $n$-dimensional matrices of natural numbers,
i.e., $\A = \mat\nats$.

Using matrix interpretations, there are at least three approaches to estimate
the value of $\semo t$ depending on $|t|$.
In all these techniques one collects the set of all matrix-coefficients 
$\MM = \{M_{f,i} \mid f \in \FF, 1 \leq i \leq \text{arity of $f$}\}$
and it is easy to see that $\semo t \leq |t| \cdot c \cdot \max \{ N_1\cdots N_{|t|} \mid N_i \in \MM\}$ where $c$ is some constant depending on $\{M_{f,0} \mid f \in \FF\}$.

\begin{enumerate}
\item 
\label{easy}
In \cite{MSW08} one approximates $\{ N_1\cdots N_{|t|} \mid N_i \in \MM\}$
  by $M_\max^{|t|}$ where $M_\max$ is the pointwise maximum matrix of $\MM$.
  Afterwards, a sufficient criterion to bound
  the value of $M_\max^{|t|}$ is provided: if $M_\max$ is upper triangular where all
  entries on the diagonal are at most 1, then the overall complexity is within
  $\OO(|t|^n)$ where $n$ is the dimension of the matrix. This result is proven
  using a standard inductive proof.
\item 
\label{cayley}
In \cite{NZM10} the previous result is extended 
  as follows:
  If all eigenvalues of the characteristic polynomial of $M_\max$ 
  are at most 1, then the overall complexity is 
  within $\OO(|t|^m)$ where $m$ is the multiplicity of eigenvalue 1.
  This result is proven via the theorem of Cayley-Hamilton \cite{Rose}.
\item 
\label{joint}
Even more precise estimations can be gained by using theorems from joint
  spectral radius theory \cite{JSR,CAI11} as these do not perform the rough 
  approximation of
  $\MM$ via $M_\max$. Unfortunately, the corresponding mathematics 
   is even
  more complicated than the Cayley-Hamilton theorem.
\end{enumerate}

\section{Formalizing Matrix Interpretations for Termination}
\label{matrix}

Recall that the aim of our formalization is to
obtain an executable program, \ceta, that is able to certify
proofs with matrix interpretations of arbitrary dimensions.

For the formalization of matrices itself, there are several options:
\begin{itemize}
\item If there are dependent types, then the obvious choice is
  to model matrices in $\mat\nats$ as lists of lists of length $n$. However,
  we are working in Isabelle/HOL, so this is not an option in our case.
\item Alternatively, one can use the idea 
  to model matrices as functions of type $I \to I \to A$ where
  $I$ is some finite index type and where the cardinality of $I$ 
  corresponds to the
  dimension \cite{Harrison05}.
  
  Unfortunately, this trick is not possible in our case, since as far as we
  see, for code-generation it is necessary to instantiate the index type $I$
  above for every dimension that we would require. However,
  the dimensions of matrices that will be used in the certificates can
  be arbitrary without any bound on the dimension.
\item The representation of Steven Obua uses a type $\nats \to \nats \to A$
  with the restriction that only finitely entries are non-zero.\footnote{See
  \texttt{HOL/Matrix-LP/Matrix.thy} for details.}
  Here, we see two problems, namely executability of matrix comparisons
  and moreover, this representation does not allow to define a 1-matrix,
  which would be inconvenient for our purposes.
\item Define matrixes just as lists of lists and use predicates as \emph{guards}
  to
  ensure that the dimensions fit.
\end{itemize}

\newcommand{\gm}[3]{\mathsf{mat}^{#1,#2}(#3)}

We formalized several matrix-operations using the last approach with guards. 
Essentially, all our theorems look as follows where $\gm mnM$ is a predicate
that ensures that $M$ represents an $m \times n$-matrix, i.e., the outer list
is of length $m$ and all inner lists have length $n$.
\begin{align}
\label{commute}
\gm mnM \Longrightarrow \gm mnN \Longrightarrow\ & M+N = N + M \\ 
\label{preserve}
\gm mnM \Longrightarrow \gm mnN \Longrightarrow\ & \gm mn{M+N} \\
\label{equiv}
\gm mnM \Longrightarrow \gm mnN \Longrightarrow\ & (M = N) = (\forall ij. M_{ij} = N_{ij}) \\
\label{plus}
\gm mnM \Longrightarrow \gm mnN \Longrightarrow\ & (M+N)_{ij} = M_{ij} + N_{ij} 
\end{align}

Here, property \rSub{commute} states that matrix addition is commutative
(silently assuming that the addition on the underlying carrier is commutative),
but this fact is guarded by the condition that the matrix dimensions of both matrices fit together.

The next kind of property \rSub{preserve} states that matrix addition preserves
the dimensions which is required to perform reasoning within contexts, e.g.,
to prove $(M + N) + K = K + (M + N)$ where we only know the dimensions of
matrices $M$, $N$, and $K$.

Property \rSub{equiv} was somehow the key to prove most properties of basic matrix
operations: Instead of comparing matrices using their representing type, i.e., the
inductive type of lists, we do a pointwise comparison. And then a property
like \rSub{plus} 
just states that the algorithm for addition (which is defined recursively over lists) is correct 
w.r.t.\ the pointwise definition of matrix addition. Afterwards, all properties
involving addition use the characterization of \rSub{plus} instead of the concrete implementation on lists.
For example, the prove of \rSub{commute} becomes trivial using \rSub{equiv},
\rSub{plus}, and commutativity of the addition on the carrier, and does not
require any induction.

Using this representation of matrices, code-generation works without any
problems, since all algorithms like matrix-addition, -multiplication, etc.\ are
just algorithms on lists.
However, it has one major disadvantage, namely that we cannot use matrices
in combination with the standard
classes like \emph{group} or \emph{semiring} from the Isabelle-distribution since
these require equalities like $M + N = N + M$ without the additional guards
that we impose. As one example consequence, it is not possible to combine the polynomial library of Clemens Ballarin
from the Isabelle-distribution\footnote{In
\texttt{HOL/Algebra/abstract/Ring2.thy} one can see the (unguarded) requirements for \texttt{HOL/Algebra/poly/Polynomial.thy}} with our matrix library.

To this end, we had to develop our own library on linear polynomials
which works on guarded operations and requires properties like \rSub{commute}
and \rSub{preserve}. It is needless to say that working with these guards
is by far more cumbersome than working with the similar unguarded classes 
from the distribution. 

%------------------------------------------------------------------------

\section{Formalizing Matrix Interpretations for Complexity}
\label{complexity}

The switch from termination to complexity proofs via matrix interpretations
poses one additional challenge, namely that of estimating values or
growth-rates of matrices. 

So far, we formalized the approach of \rSub{easy}
using triangular matrices. 
Already in this technique, an unexpected challenge has occurred in formalizing that 
the linear matrix norm is sub-multiplicative, i.e., $\norm{M \cdot N} \leq
\norm M \cdot \norm N$. In the literature we only found proofs for matrices
over real or
complex numbers which are based on a suprema- or limit-construction. However, in our setting
we would like to have this statement for matrices over arbitrary carriers
like the
naturals, the integers, or the rationals. Therefore, we developed our own proof
which works by induction over the shared dimension of $M$ and $N$ and is about
200 lines long (in Isabelle).

For the future, when extending our work towards the more sophisticated methods
of \rSub{cayley} and \rSub{joint} we would like to minimize the effort in finding
new proofs, e.g., for \rSub{cayley} we plan to first formalize the theorem
of Cayley-Hamilton and then use it in the same way as it is done in \cite{NZM10}.
 However, here already in the setup there is one major obstacle:
the theorem of Cayley-Hamilton requires non-linear polynomials over
  matrices, and we are not aware of any Isabelle library on 
  non-linear polynomials
  that is able to deal with guarded semirings like our matrices.
  
So, the questions to the Isabelle-community would be, whether
\begin{itemize}
\item someone has already done work on non-linear polynomials using guarded semirings?
\item one should try to generalize the existing classes like semiring 
  and the existing polynomial library from the
  distribution to work with guards?
\item one should develop an independent formalization
  of non-linear polynomials including guards?
\item we overlooked something, and there is a possibility to use matrices
  of arbitrary dimension in combination with code-extraction and 
  the existing library on
  polynomials?
\end{itemize}

\bibliographystyle{plain}
\bibliography{complexity}

\end{document}